\documentclass[preprint,12pt,authoryear]{elsarticle}

\usepackage{amssymb}
\usepackage{amsmath}
\usepackage{color}
\usepackage{multirow}
\usepackage{subfigure}
\usepackage{makecell}

\journal{Elsevier journal}

\begin{document}
	
	\begin{frontmatter}
		
		
		
		\title{Pay Less On Clinical Images: Asymmetric Multi-Modal Fusion Method For Efficient Multi-Label Skin Lesion Classification}

		\author[label1,label2]{Peng TANG}
%		\author[label3]{Xintong YAN}
%		\author[label4]{Yang NAN}
%		\author[label1,label5]{Xiaobin HU}
%		\author[label1,label5]{Xiaobin HU}
%		\author[label5]{Bjoern H. Menze}
%		\author[label6]{Sebastian Krammer}
		\author[label1,label2]{Tobias Lasser}

		\address[label1]{organization={Department of Informatics, School of Computation, Information, and Technology, Technical University of Munich},%Department and Organization
			%addressline={}, 
			city={Garching},
			%postcode={85748}, 
			%state={},
			country={Germany},
			e-mail={tangp@in.tum.de}}
		
		\address[label2]{organization={Munich Institute of Biomedical Engineering, Technical University of Munich},%Department and Organization
			%addressline={}, 
			city={Garching},
			%postcode={85748}, 
			%state={},
			country={Germany},
			e-mail={lasser@cit.tum.de}
		}

		\begin{abstract}
		Existing multi-modal approaches primarily focus on enhancing multi-label skin lesion classification performance through advanced fusion modules, often neglecting the associated rise in parameters. 
		In clinical settings, both clinical and dermoscopy images are captured for diagnosis; however, dermoscopy images exhibit more crucial visual features for multi-label skin lesion classification.
		Motivated by this observation, we introduce a novel asymmetric multi-modal fusion method in this paper for efficient multi-label skin lesion classification.
		Our fusion method incorporates two innovative schemes.
		Firstly, we validate the effectiveness of our asymmetric fusion structure. It employs a light and simple network for clinical images and a heavier, more complex one for dermoscopy images, resulting in significant parameter savings compared to the symmetric fusion structure using two identical networks for both modalities.
		Secondly, in contrast to previous approaches using mutual attention modules for interaction between image modalities, we propose an asymmetric attention module. This module solely leverages clinical image information to enhance dermoscopy image features, considering clinical images as supplementary information in our pipeline.
		We conduct the extensive experiments on the seven-point checklist dataset.
		Results demonstrate the generality of our proposed method for both networks and Transformer structures, showcasing its superiority over existing methods
		We will make our code publicly available.
		\end{abstract}
		
	%	\begin{graphicalabstract}
	%	\end{graphicalabstract}
		
	%	\begin{highlights}
	%		\item Research highlight 1
	%		\item Research highlight 2
%		\end{highlights}
		
		\begin{keyword}
		skin lesion classification,	multi-modal learning, asymmetric Fusion, Clinical and Dermoscopy Images
			
			
			
		\end{keyword}
		
	\end{frontmatter}
	
	

\begin{figure*}[h]  %子图加并列
	\centering
	\subfigure[CE=DE]{
		\begin{minipage}[h]{0.47\textwidth}
			\includegraphics[width=6cm,height=3.5cm]{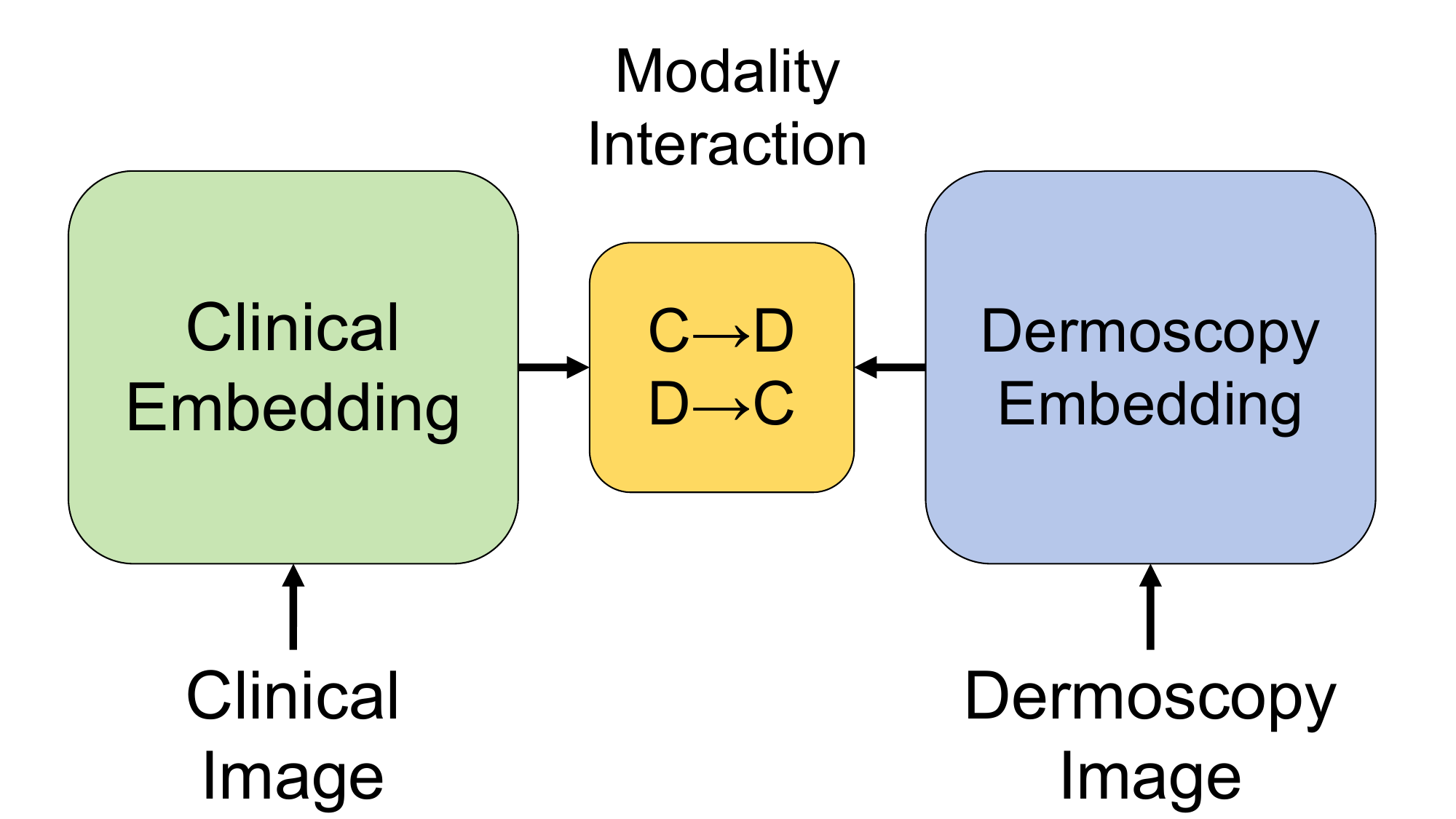} 
		\end{minipage}
	}
	\subfigure[CE$\textless$DE]{
		\begin{minipage}[h]{0.47\textwidth}
			\includegraphics[width=6cm,height=3.5cm]{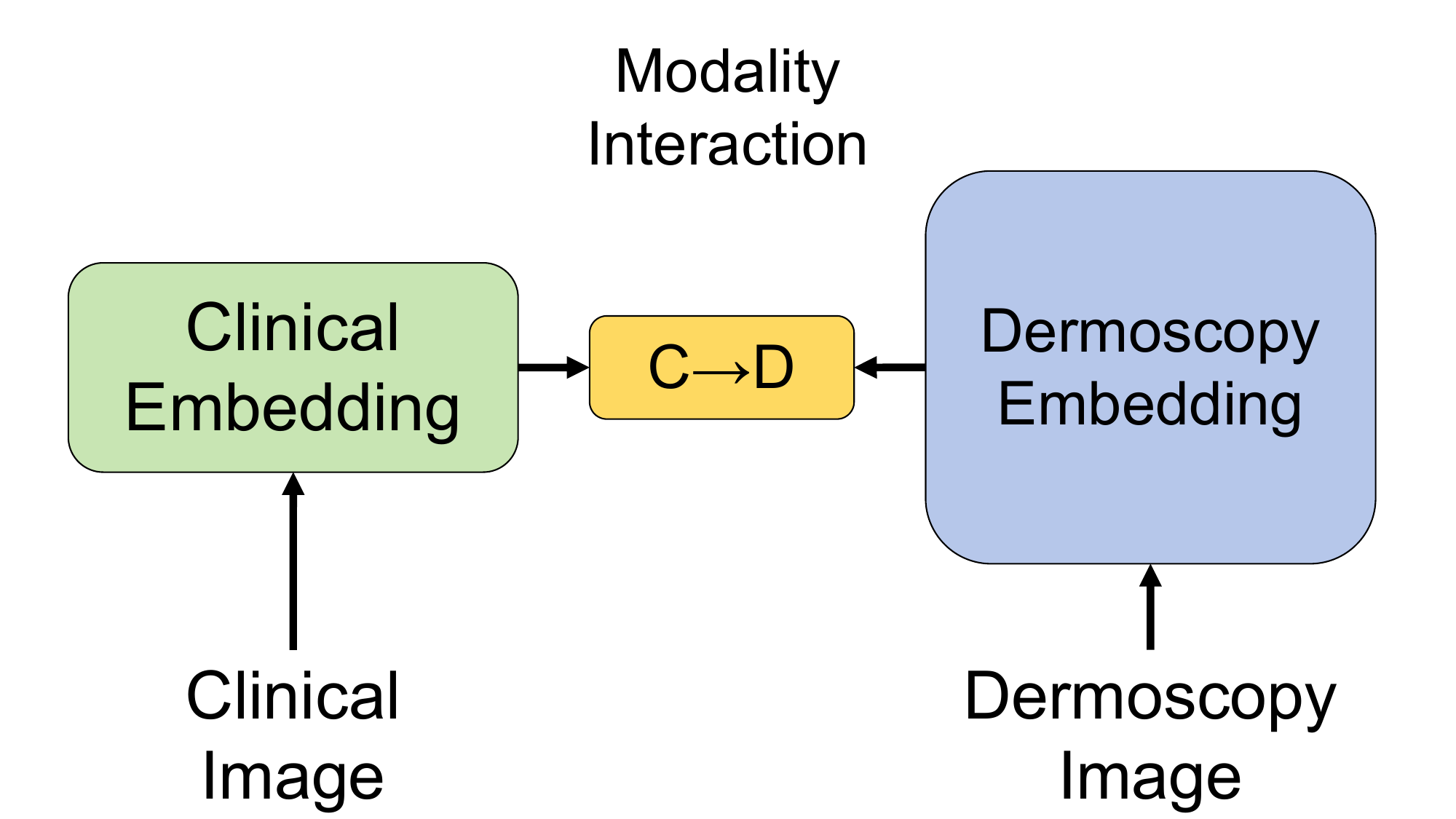} 
			
		\end{minipage}
	}
	\caption{The comparison between (a) state-of-the-art multi-modal skin lesion classification methods and (b) our proposed method. The height of each rectangle indicates its relative computational size. CE and DE are shorthand for clinical embedding and dermoscopy embedding, respectively.}
	\label{fig1}
	\vspace{-0.2in}
\end{figure*}

\section{Introduction}
\label{sec1}
Early detection of melanoma holds crucial significance for patient treatment.
Patients with ulcerated melanoma exceeding 4 mm thickness experience a 15$\%$ five-year survival rate, whereas those with the melanoma thinner than 1 mm exhibit a 95$\%$ 5-year survival rate \citep{balch2009final, vestergaard2008dermoscopy}.
However, early-stage melanoma prevention faces limitations due to the restricted count of experienced dermatologists. 
It is anticipated that deep learning-based decision support systems will enhance the diagnostic accuracy of young dermatologists and may even serve as a potential replacement for experienced dermatologists \citep{zhang2023tformer}.

Mimicking routine dermatologist's examinations the seven-point checklist (SPC) dataset was introduced using clinical screening followed by dermoscopic analysis \citep{ge2017skin}, \cite{kawahara2018seven}. 
This dataset is comprised of paired clinical-dermoscopy images, facilitating research in multi-modal skin lesion classification (MM-SLC).
For the presentation of localized visual features, dermoscopy images (DIs) are captured using a high-resolution magnifying imaging device \citep{vestergaard2008dermoscopy} in direct contact with the skin (e.g. dermatoscopy and epiluminescence microscopy). 
In contrast, clinical images (CIs), taken with a standard digital camera or smartphone, exhibit more variations in terms of perspective and angle \citep{ge2017skin}.
In contrast to single modality-based SLC, MM-SLC harnesses complementary information from both modalities and leads to a more accurate and robust diagnosis, driving further exploration on this topic \citep{zhang2023tformer}.

%% Introduction of multi-modal skin lesion analysis methods
Current MM-SLC approaches \citep{ge2017skin, yap2018multimodal, kawahara2018seven, bi2020multi, tang2022fusionm4net, fu2022graph,  he2023co, zhang2023tformer} are predominantly based on deep learning structures, i.e., convolution neural networks (CNN) and transformers (TF). 
All the aforementioned approaches utilize a symmetric multi-modal framework, as depicted in Figure \ref{fig1}(a), where features from both image modalities are extracted using identical deep learning models and are subsequently fused through a modality interaction operation.
Compared to simple concatenation \citep{yap2018multimodal, ge2017skin, fu2022graph} and summation \citep{tang2022fusionm4net}, recent works have increasingly utilized mutual modality attention mechanisms \citep{bi2020multi, he2023co, zhang2023tformer}, achieving higher accuracy.
However, emphasizing the supplementary information of Clinical Images (CI) in the mutual-attention mechanism may impact diagnostic results. 
Furthermore, all these methods require a relatively large number of model parameters with associated high computational costs, limiting their practical use in clinics. 
For example, in some rural areas, medical resources are scarce. While tele-dermatology can extend access to specialists online \citep{liu2020deep}, poor infrastructure may not guarantee reliable network signals for remote diagnosis. In such cases, the local deployment of AI algorithms becomes a better choice. Reducing the parameters of these algorithms, including SkinGPT-4 \citep{zhou2023skingpt} which has demonstrated potential for 24-hour care of skin diseases, contributes to cheaper cost and thus enables broader usage. 

Our primary objective is to design a MM-SLC framework that achieves a favorable parameter/accuracy trade-off, i.e., substantially reducing the model's parameters while only modestly impacting its accuracy.
Our idea is motivated by two observations: 
(1) Clinical Statistics: According to experienced dermatologists, the diagnostic accuracy of melanoma based on Dermoscopy Imaging (DI) is 25$\%$ higher than that of visual inspection with naked eyes. 
Visual features observed by naked eyes are akin to those captured by standard cameras and smartphones \citep{togawa2023comparison}. 
(2) Experimental Results of DL Algorithms: In diagnosis tasks using the SPC dataset, the majority of papers \citep{kawahara2018seven, bi2020multi, tang2022fusionm4net, fu2022graph, he2023co, zhang2023tformer} reported a 6$\%$ higher accuracy in Dermoscopy Imaging (DI)-based diagnosis compared to Clinical Image (CI). 
Additionally, \citep{dascalu2022non} demonstrated an increased diagnostic accuracy with DI (85$\%$) compared to CI (75$\%$).
Considering both observations, it is evident that a significant amount of key diagnostic information comes from DI rather than CI.
Therefore, employing two identical structures to extract information from DI and CI individually appears to be wasteful.

Inspired by that, in this paper, we propose a novel Asymmetrical Multi-Modal Fusion Method (AMMFM) for efficient multi-label skin lesion classification.
Our approach differentiates itself from previous methods in two key aspects, i.e., Asymmetric Fusion Framework (AFF) and Asymmetric Attention Block (AAB).
First, differing from the commonly-used Symmetrical Fusion Framework (SFF), our AFF incorporates the prior domain knowledge into the structure design. 
AFF utilizes an advanced model, e.g., ResNet, ConvNext, and SwinTransformer \citep{he2016deep, liu2022convnet, liu2022video} for capturing the primary diagnostic information from DI, but a lightweight deep model, i.e., MobilenetV3 \citep{howard2019searching}, for the supplementary information from CI. 
Compared to SFF, AFF significantly reduces the model's parameters with only a subtle decrease in accuracy.
Second, in contrast to previous methods utilizing bidirectional attention blocks (BAB) to mutually enhance DI and CI (Fig.~\ref{fig1}.a), we believe that enhancing the supplementary information CI may lead to overfitting, affecting the final classification. 
Therefore, we propose an asymmetric attention block (AAB) to exclusively leverage the features of CI to enhance those of DI (Fig.~\ref{fig1}.b), achieving superior performance compared to BAB with fewer model parameters.
In total, our contributions can be summarized as follows:

\begin{enumerate}
	\item Inspired by prior knowledge, we introduce a novel  Asymmetrical Fusion Framework (AFF) that significantly reduces the model's parameters while maintaining unchanged or slightly decreased classification accuracy compared to the currently used Symmetric Fusion Framework (SFF).
	
	\item We present a new Asymmetrical Attention Block (AAB) that exclusively utilizes features extracted from clinical images (CI) to enhance those of dermoscopy images. This approach addresses potential accuracy impacts associated with focusing on supplementary information from CI. In comparison to the former Bidirectional Attention Block (BAB), our AAB demonstrates improved classification performance with fewer parameters.
	
	\item Our proposed Asymmetrical Multi-Modal Fusion Method achieves state-of-the-art performance in both accuracy and model's parameters. The extensive results confirm the effectiveness of our proposed AFF and AAB, demonstrating their applicability to various deep learning algorithms, including both CNN and transformer structures.
	
\end{enumerate}

\section{Related works}
%Single Modality
\subsection{Single-modality based skin lesion classification}
Prior to the release of the first public multi-modal skin lesion classification dataset, SPC, by \cite{kawahara2018seven}, most researchers developed their skin lesion classification (SLC) methods exclusively using dermoscopy images (DI). 
Current SLC methods based on DI \citep{Yu2020skin, tang2020gp, liu2022ci, sarker2022transslc, yang2023novel} predominantly rely on CNN or transformer structures, leveraging the success of these algorithms in computer vision tasks.
For example, \cite{Yu2020skin} secured the first place in the ISBI-2016 SLC challenge by proposing a deep residual CNN with the assistance of a segmentation model. 
Additionally, \cite{yang2023novel} introduced a novel vision transformer for SLC, surpassing state-of-the-art methods on the HAM10000 dataset \citep{tschandl2018ham10000}.

\subsection{Multi-modality based skin lesion classification}
Despite their success, these methods tend to deviate from routine dermatologists' examinations \cite{zhang2023tformer} and overlook the potential to enhance diagnostic accuracy by exploiting complementary information from additional modalities. 
To fill this gap, several works about MM-SLC were presented \citep{ge2017skin, yap2018multimodal, kawahara2018seven, tang2022fusionm4net, fu2022graph, he2023co, zhang2023tformer}. 
\cite{ge2017skin} and \cite{yap2018multimodal} extracted the features from clinical images and dermoscopy images using VGG-16 \citep{simonyan2014very} and Resnet-50 \citep{he2016deep}, and then fused the features by a simple concatenation to learn a joint representation for the final prediction.
The introduction of the Seven-Point Checklist (SPC) dataset by \cite{kawahara2018seven} marked a significant advancement in multi-modal skin lesion classification. 
They proposed a multi-modal framework based on Inception-V3 \citep{szegedy2016rethinking} for the simultaneous classification of diagnosis and the seven-point checklist.
After the release of the SPC dataset, there has been a growing number of multi-modal approaches proposed for multi-label skin lesion classification.
To enhance performance, \cite{tang2022fusionm4net} and \cite{fu2022graph} introduced weighted-fusion and graph-based fusion schemes, respectively. 
Both approaches combine predictions from CI and DI in the late stages of the model.
More recently, \cite{he2023co} and \cite{zhang2023tformer} recognized the limitations of the simple concatenation operation used in former methods. 
They introduced multiple bidirectional attention blocks to mutually enhance CI and DI, facilitating efficient interaction between these two modalities across multiple scales.
While these methods have achieved state-of-the-art (SOTA) performance for Multi-Modal Skin Lesion Classification (MM-SLC) tasks, the emphasis on Clinical Images (CI) may potentially impact their performance, considering CI is regarded as supplementary information in our opinion. 
Additionally, the substantial number of parameters and associated high computational cost in these methods may pose limitations in various clinical scenarios.

\subsection{Asymmetrical Fusion Method}
A limited number of works on the Asymmetrical Fusion Model (AFM) \citep{yang2019asymmetric, zhu2019asymmetric, gao2020asymmetric, yang2021asymmetric, wu2023asymmetric} have been proposed for various computer vision tasks.
For example, \cite{zhu2019asymmetric} introduced an asymmetric non-local network to fuse multi-scale features for semantic segmentation. 
\cite{gao2020asymmetric} presented an asymmetrical model to extract asymmetric relations between humans and objectives for action recognition.  
Additionally, \cite{wu2023asymmetric} proposed an asymmetrical fusion framework with a focus on the gallery side for image retrieval.
In the medical domain, \cite{yang2023novel} employed an asymmetrical model to address issues in 3D slices for universal lesion detection.

These methods are tailored to specific modalities (e.g., 3D slices \citep{gao2020asymmetric}) or tasks (e.g., semantic segmentation \citep{zhu2019asymmetric}, action recognition \citep{gao2020asymmetric}, and image-text retrieval \citep{wu2023asymmetric}). 
Their design is not directly applicable to the Multi-Modal Skin Lesion Classification (MM-SLC) task, however, the successes of AFM in different fields encouraged us to explore the potential of asymmetrical fusion methods for the MM-SLC task.

\begin{figure*}
	\centering
	\includegraphics[width=14cm,height=7cm]{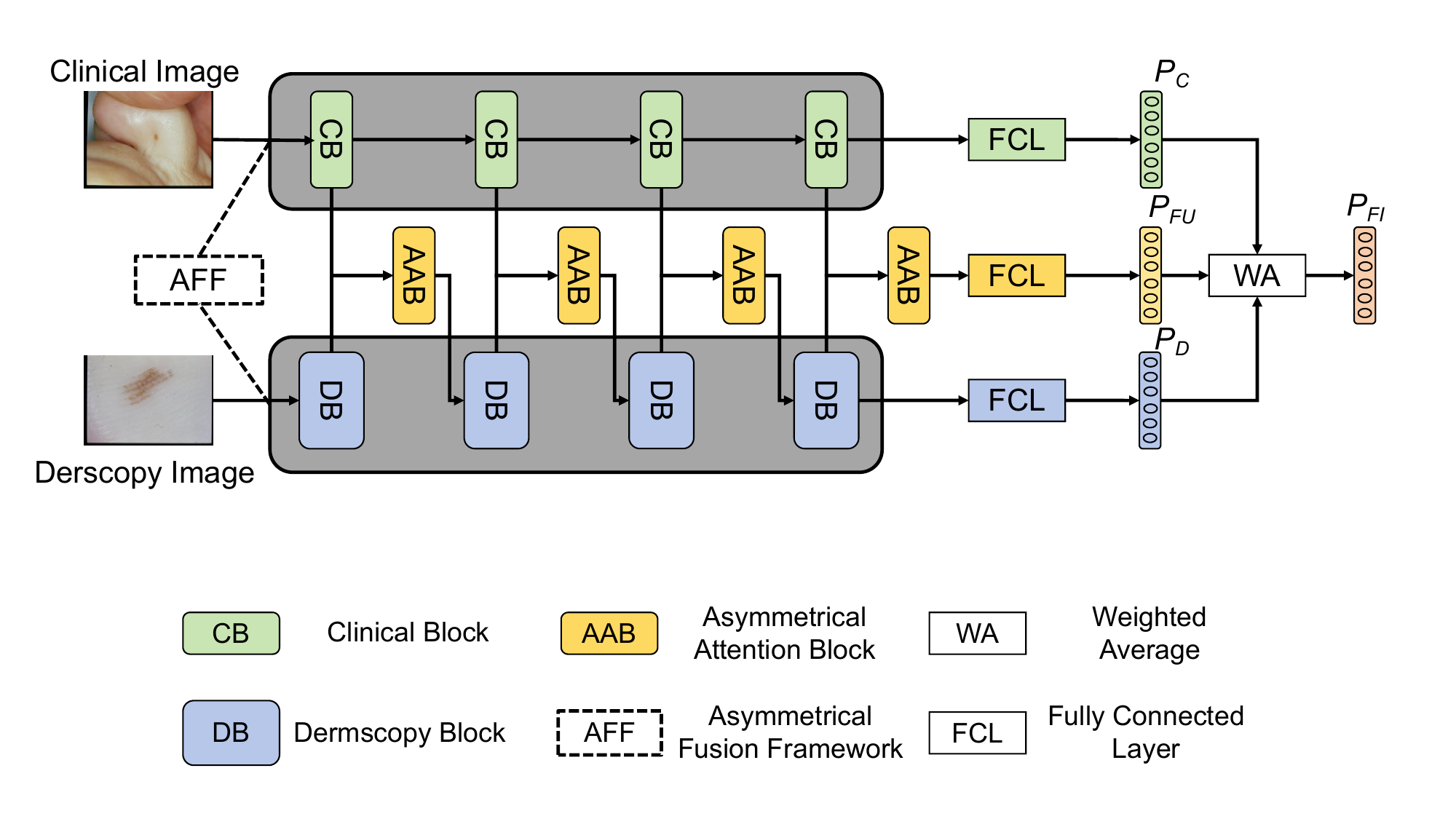}
	\caption{The overview of our Asymmetric Multi-Modal Fusion Method. Clinical and dermoscopy blocks are used to extract the features from clinical and dermoscopy image, respectively. $P_C$, $P_D$ and $P_{FU}$ are the predictions from clinical branch (green), dermoscopy branch (blue) and fusion branch (yellow). }
	\label{fig2}
	\vspace{-0.2in}
\end{figure*}

\section{Asymmetrical Multi-Modal Fusion Method}
As illustrated in Fig.~\ref{fig2}, the proposed Asymmetric Multi-Modal Fusion Method (AMMFM) is constructed with four components: Asymmetric Fusion Framework (AFF), Asymmetric Attention Blocks (AABs), Fully Connected Layers (FCLs), and a Weighted Averaging (WA) operation.
AFF and AABs are pivotal components used to extract information from two modalities and facilitate modality interactions. 
These two components are the core to our methodology, and we will delve into their details in the following subsections.
FCLs are used to classify the extracted features in three branches: clinical branch (highlighted in green), dermoscopy branch (highlighted in blue), and fusion branch (highlighted in yellow) as depicted in Fig.~\ref{fig2}.
WA is employed to fuse predictions from clinical, dermosocopy and fusion branches, namely, $P_C$, $P_D$, and $P_{FU}$, resulting in the final prediction $P_{FI}$.

\begin{figure*}[h]  %子图加并列
	\centering
	\subfigure[Bidirectional Attention Block (BAB)]{
		\begin{minipage}[h]{0.47\textwidth}
			\includegraphics[width=7.5cm,height=4.7cm]{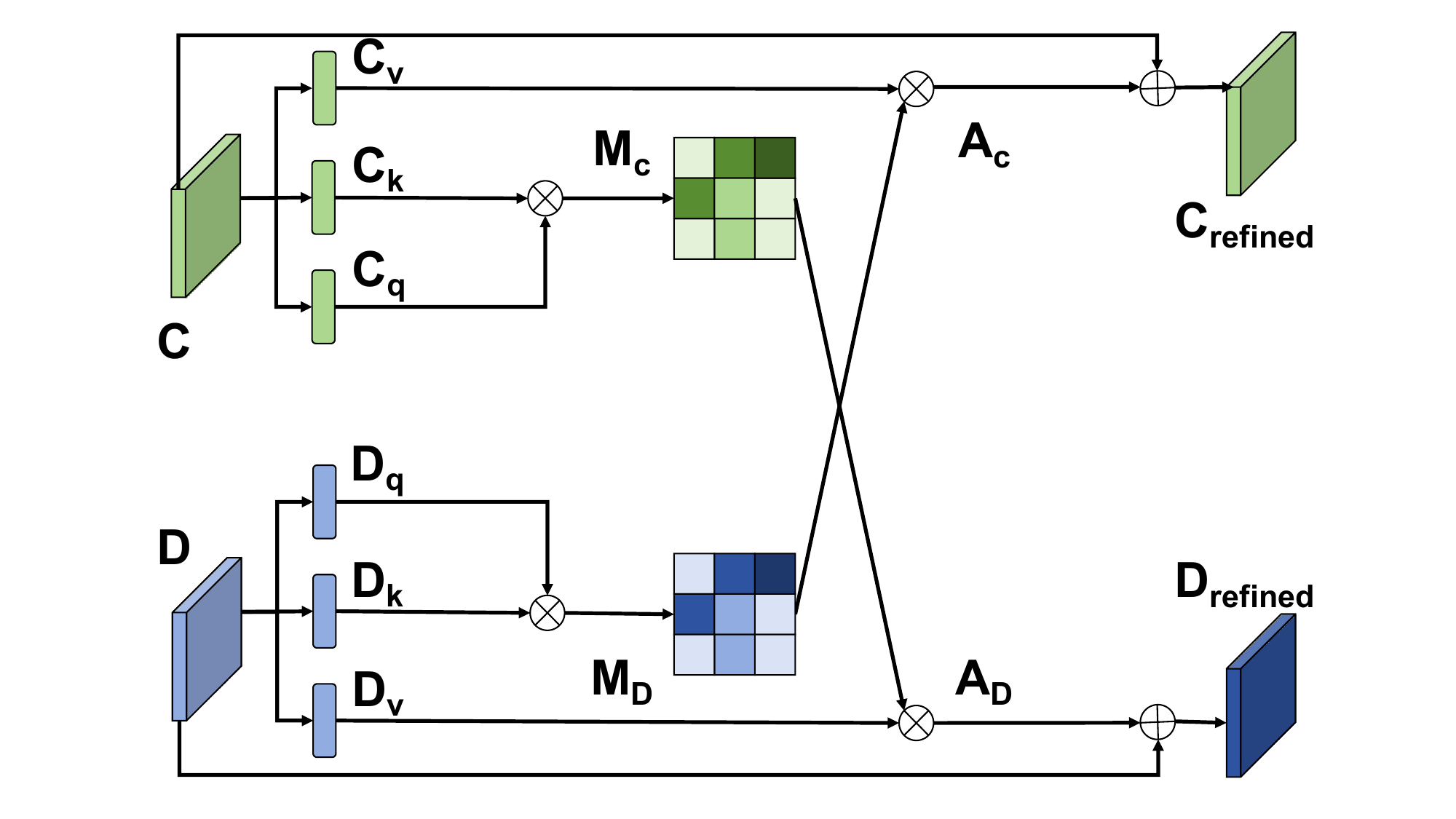} 
		\end{minipage}
	}
	\subfigure[Asymmetrical Attention Block (AAB)]{
		\begin{minipage}[h]{0.47\textwidth}
			\includegraphics[width=7.5cm,height=4.7cm]{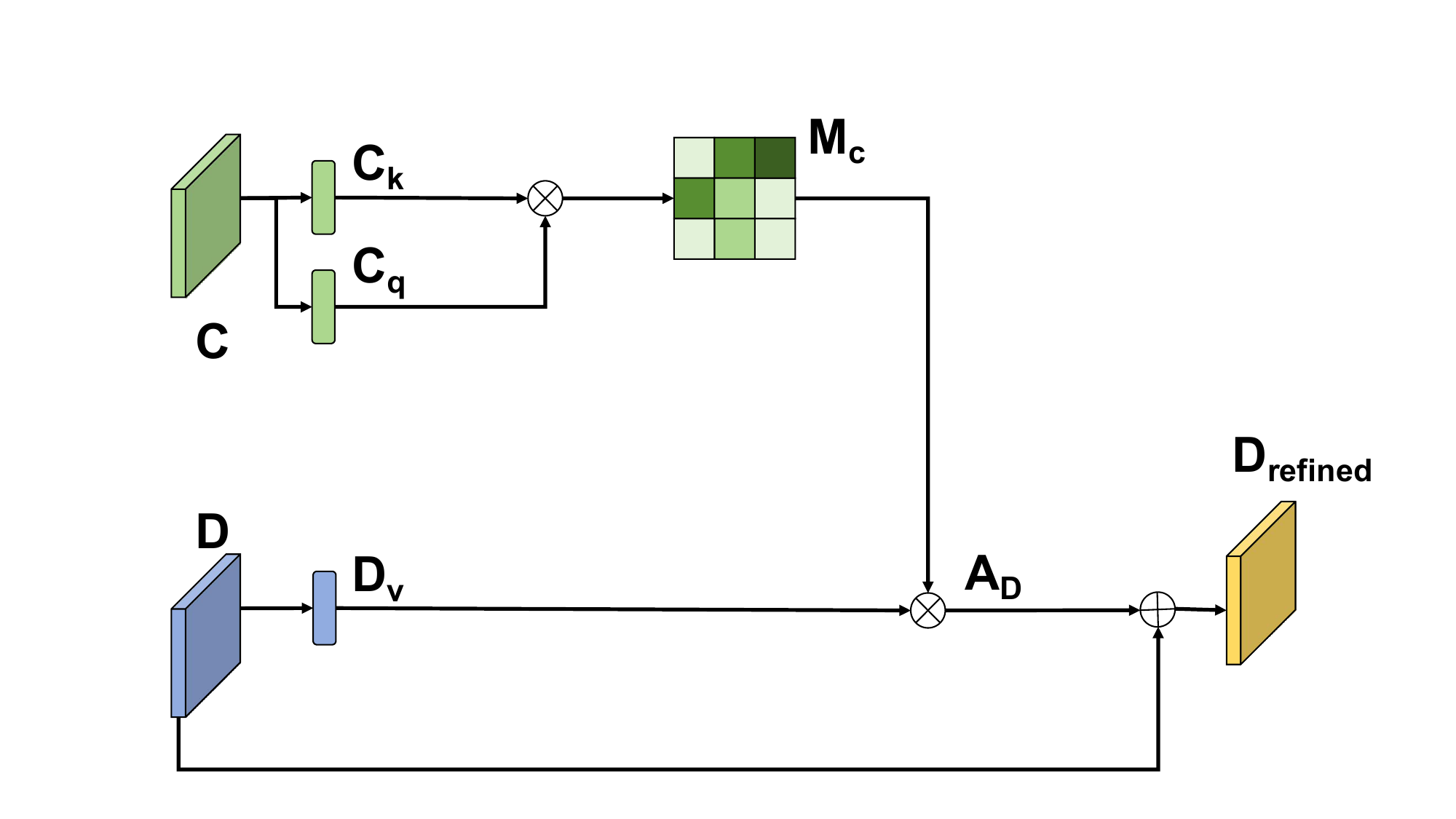} 
			
		\end{minipage}
	}
	\caption{The computational graph of a state-of-the-art bidirectional attention block (BAB) and our asymmetrical attention block (AAB).}
	\label{fig3}
	\vspace{-0.2in}
\end{figure*}

\subsection{Asymmetric Fusion Framework}
Before introducing the Asymmetric Fusion Framework (AFF), we provide a definition for easier understanding: in this article, the term ``framework'' specifically refers to the composition of two feature extractors, excluding the modality interaction modules (see Fig.~\ref{fig2}).

As discussed in Sec.~\ref{sec1}, the majority of current methods are based on a symmetrical fusion structure (SFF), utilizing two identical structures to extract features from clinical and dermoscopy images, respectively. 
However, relevant research has demonstrated that the accuracy based on dermoscopy is much higher than that based on naked eyes and clinical images captured by smartphones with a standard digital camera. 
Additionally, the seven-point checklist criteria are proposed based on features observed under dermoscopy. These phenomena drive us to consider dermoscopy images as the primary, with clinical images regarded as supplementary for the multi-label skin lesion classification task. 
Therefore, we propose an asymmetrical fusion structure (AFF) for this task.

Specifically, as shown in Fig.~\ref{fig2}, AFF utilizes two different models for the feature extractions from two modalities: a lightweight model, MobilenetV3 \citep{howard2019searching}, for clinical images (CI) and an advanced model requiring much more parameters, e.g., ResNet, ConvNext, or SwinTransformer \citep{he2016deep, liu2022convnet, liu2022video} for dermoscopy images (DI).
We believe that, compared to SFF, our AFF can significantly reduce the model's parameters while maintaining unchanged or subtly decreased classification performance. 
While replacing an advanced model with a lightweight model in the clinical branch may affect the information capture from Clinical Images (CI), it is important to note that clinical images are considered supplementary in our pipeline. 
Therefore, this change is expected to have only a slight impact on the final classification.

\subsection{Asymmetric Attention Block}
Building on the discussion in Sec.~\ref{sec1}, the information from the clinical branch is considered as supplementary in this paper. 
% Treating it as equal to dermoscopy information and enhancing the supplementary information are not reasonable in our pipeline.
Therefore, we introduce an asymmetric attention block (AAB) for the modality interactions between clinical and dermoscopy images.

In contrast to the state-of-the-art bidirectional attention block (BAB) \cite{he2023co} that mutually enhances the features of both modalities (see Fig.\ref{fig3}(a)), our AAB only adopts clinical features to generate an attention map for enhancing dermoscopy features. 
This design allows us to save approximately half of the parameters compared to BAB (See Fig.\ref{fig3}(b)).
Similar to BAB, AAB is embedded into different stages of deep learning models to facilitate the interaction of multi-scale features from the two modalities.

In our AAB, the inputs are the extracted clinical features $C \in \mathbb{R}^{H \times W \times C}$ and dermoscopy features $D \in \mathbb{R}^{H \times W \times C}$, both of which have the same size ($H$, $W$, and $C$ indicate the height, width, and channel number of the features, respectively).
As shown in Fig.~\ref{fig3}(a), firstly, two 1$\times$1 convolutions are applied to $C$ to generate $C_k$ and $C_q$, and one 1$\times$1 convolution is employed on $D$ to obtain $D_v$, where $(C_k, C_q, D_v) \in \mathbb{R}^{H \times W \times C}$. Then, $C_k$ and $C_q$ are reshaped to $\mathbb{R}^{N \times C}$, where $N = H \times W$. Next, a multiplication is conducted between the reshaped $C_k$ and $C_q$, followed by a non-linear activation function $softmax$ to generate the attention map $M_c \in \mathbb{R}^{N \times N}$.
Finally, the refined dermscopy features are obtained based on Eq.~\ref{eq1},
\begin{equation}
	D_{refined} = D_v \cdot M_c + D,
	\label{eq1}
\end{equation}
where $\cdot$ indicates matrix dot product operation, and $+$ indicates matrix summation.

\subsection{Loss Function and Final Prediction}
The total loss $L_{total}$ used to optimize our model is as follows:
\begin{equation}
	L_{total} = L_{derm} + L_{clic} + L_{fusion},
	\label{eq2}
\end{equation}
where $L_{derm}$, $L_{clic}$ and $L_{fusion}$ are the multi-label classification losses for the dermscopy image branch ($P_{D}$ in Fig.~\ref{fig2}), clinical image branch ($P_{C}$ in Fig.~\ref{fig2}) and fusion image branch ($P_{FU}$ in Fig.~\ref{fig2}), respectively.
All the losses are computed as 
\begin{equation}
	L_{K}=\sum_j^X \sum_i^Y CE\left(D^j, C^j, G_i^j, P_i^j ; \theta_K\right),
	\label{eq3}
\end{equation}
where $X$ is the batch size in our training, $Y$=8 indicates the number of the multi-label classification tasks (see Table~\ref{table1}), $C^{j}$ and $D^{j}$ represent input pairs of dermoscopy and clinical images, respectively. $G_i^j$ and $P_i^j$ are the corresponding ground truths and predictions, respectively, and $\theta_{K}$ is the parameters of our model.  
$CE$ indicates the cross-entropy loss.

During the testing stage, we use a weighted average scheme to fuse $P_{D}$, $P_{C}$ and $P_{FU}$ into the final prediction $P_{FI}$ for the evaluation as follows:
\begin{equation}
	P_{FI} = W_{D} * P_{D} + W_{C} * P_{C} + W_{FU} * P_{FU},
	\label{eq4}
\end{equation}
where $W_D$, $W_C$ and $W_{FU}$ are the corresponding weights for $P_D$, $P_C$ and $P_{FU}$, respectively, which are obtained by the conducting a weight search scheme on the validation dataset \citep{tang2022fusionm4net}.

\section{Experiments and Discussion}
\subsection{Implementation Details}
During training, we use Adam \citep{kingma2014adam} with a batch size of 24 to optimize our model for 250 epochs. 
Data augmentations, including flipping, shifting, scaling, rotating, and brightening operations, are randomly conducted to enhance the generalization ability of the model. 
The Stochastic Weights Averaging (SWA) \cite{izmailov2018averaging} scheme is used in the last 50 epochs to generate the final weights for evaluation. 
All images are resized to $224 \times 224 \times 3$ for both training and evaluating the model. 
Following \cite{tang2022fusionm4net}, test time augmentation is also used during the evaluation to improve the classification performance. 
All of our experiments are conducted on NVIDIA GPUs A100 (40GB). 
Unless otherwise specified, our AMMFM are based on MobilenetV3 for clinical images and Swin-Transformer for dermoscopy images, as it achieves the best classification performance in our experiments.  
More details can be found in our released code.

\subsection{Dataset and Metrics}
The effectiveness of our AMMFM is evaluated on the well-recognized seven-point checklist (SPC) dataset \cite{kawahara2018seven}, which contains 1011 patient cases. 
Each case includes a pair of dermoscopy and clinical images, diagnosis (Diag) label, and labels of seven-point checklist (SPC) features. 
As shown in Table \ref{table1}, Diag has five categories: BCC, NEV, MEL, MISC, and SK.
The SPC labels include Pigment Network (PN), Streaks (STR), Pigmentation (PIG), Regression Structures (RS), Dots and Globules (DaG), Blue Whitish Veil (BWV), and Vascular Structures (VS), which are divided into the following categories: ABS, PRS, TYP, ATP, REG, and IR.

Building on previous works \cite{kawahara2018seven, fu2022graph, tang2022fusionm4net, he2023co, zhang2023tformer}, our evaluation metrics include accuracy (ACC), area under the curve (AUC), precision (Prec), specificity (SPE), and sensitivity (SEN) for evaluation of our method.

% \usepackage{multirow}
\begin{table}[t]
	\centering
	\caption{Details of the SPC dataset.}
	\begin{tabular}{cccc}
		\hline
		Classification Task   & Name                 & Abbrev. & Num. \\ \hline
		\multirow{5}{*}{Diag} & Basal Cell Carcinoma & BCC     & 42   \\
		& Nevus                & NEV     & 575  \\
		& Melanoma             & MEL     & 252  \\
		& Miscellanoeous       & MISC    & 97   \\
		& Seborrheic Keratosis & SK      & 45   \\ \hline
		\multirow{3}{*}{PN}   & Absent               & ABS     & 400  \\
		& Typical              & TYP     & 381  \\
		& Atypical             & ATP     & 230  \\ \hline
		\multirow{3}{*}{STR}  & Absent               & ABS     & 653  \\
		& Regular              & REG     & 107  \\
		& Irregular            & IR      & 251  \\ \hline
		\multirow{3}{*}{PIG}  & Absent               & ABS     & 588  \\
		& Regular              & REG     & 118  \\
		& Irregular            & IR      & 305  \\ \hline
		\multirow{2}{*}{RS}   & Absent               & ABS     & 758  \\
		& Present              & PRS     & 253  \\ \hline
		\multirow{3}{*}{DaG}  & Absent               & ABS     & 229  \\
		& Regular              & REG     & 334  \\
		& Irregular            & IR      & 448  \\ \hline
		\multirow{2}{*}{BWV}  & Absent               & ABS     & 816  \\
		& Present              & PRS     & 195  \\ \hline
		\multirow{3}{*}{VS}   & Absent               & ABS     & 833  \\
		& Regular              & REG     & 117  \\
		& Irregular            & IR      & 71   \\ \hline
	\end{tabular}
	\label{table1}
	\vspace{-0.2in}
\end{table}
% \usepackage{multirow}

\begin{table*}[t]
	\centering
	\caption{Comparison between our proposed method AMMFM and other state-of-the-art methods based on averaged AUC values. 
		The highest and second highest values in each column are marked in bold and italics, respectively. Incep-com: Inception-combined, FM-FS: FusionM4Net-FS, AVG: Averaged ($\%$)}
	\centering
	\tabcolsep=1mm
	\renewcommand\arraystretch{1.2}
	\scalebox{0.65}{
		\begin{tabular}{c|cccccccccccccccccc}
			\hline
			\multirow{2}{*}{Methods} & \multicolumn{5}{c}{Diag}                                                      & \multicolumn{2}{c}{PN}        & \multicolumn{2}{c}{STR}       & \multicolumn{2}{c}{PIG}       & RS            & \multicolumn{2}{c}{DAG}       & BWV           & \multicolumn{2}{c}{VS}        & \multirow{2}{*}{\color{red}{AVG}} \\ \cline{2-18}
			& BCC           & NEV           & MEL           & MISC          & SK            & TYP           & ATP           & REG           & IR            & REG           & IR            & PRS           & REG           & IR            & PRS           & REG           & IR            &                      \\ \hline
			Incep-com            & 92.9          & 89.7          & 86.3          & 88.3          & 91            & 84.2          & 79.9          & 87            & 78.9          & 74.9          & 79            & 82.9          & 76.5          & 79.9          & 89.2          & 85.5          & 76.1          & 83.7                 \\
			HcCNN                    & 94.4          & 87.7          & 85.6          & 88.3          & 80.4          & 85.9          & 78.3          & 87.8          & 77.6          & \textbf{83.6} & 81.3          & 81.9          & 77.7          & 82.6          & 89.8          & 87            & \textit{82.7} & 84.3                 \\
			FM-FS                    & 95.3          & \textit{92.6} & 89            & \textbf{94}   & 89.2          & \textit{85.9} & 83.9          & \textit{87.9} & 81.4          & 80.9          & 83.5          & 81.7          & \textit{79.1} & 80.1          & 90.6          & \textit{87.8} & 78            & 86                   \\
			GIIN                     & 92.8          & 86.8          & 87.6          & 88.8          & 79.8          & 80.1          & \textbf{87.5} & 84.9          & 81.2          & 81.1          & 83.6          & 79            & 78.6          & \textit{83.1} & 90.8          & 80.7          & 75.4          & 83.6                 \\
			CAFNet                   & \textbf{97.1} & \textbf{92.7} & \textbf{92.2} & 92.5          & \textit{91}   & 81.9          & 75.3          & 87.4          & \textbf{85.4} & 76.1          & \textit{85}   & \textbf{85.4} & 75.2          & 78.7          & \textbf{94.7} & 84.8          & \textbf{83.5} & 85.8                 \\
			\textbf{AMMFM}                    & \textit{95.8} & 92.4          & \textit{89.3} & \textit{93.7} & \textbf{91.7} & \textbf{87.1} & \textit{86.7} & \textbf{91.5} & \textit{85.2} & \textit{82.5} & \textbf{86.2} & \textit{83.6} & \textbf{79.8} & \textbf{86.0} & \textit{94.1} & \textbf{90.2} & 82.4          & \textbf{88.1}        \\ \hline
	\end{tabular}}
	\label{tab2}
	\vspace{-0.1in}
\end{table*}

\begin{table}[t]
	\centering
	\caption{Comparison between our proposed method AMMFM and other state-of-the-art methods based on averaged ACC values. 
		The highest and second highest values in each column are marked in bold and italics, respectively. Incep-com: Inception-combined, FM-FS: FusionM4Net-FS, AVG: Averaged ($\%$)}
	\tabcolsep=1mm
	\renewcommand\arraystretch{1.2}
	\scalebox{0.8}{
		\begin{tabular}{cccccccccc}
			\hline
			Methods        & PN            & BWV           & VS            & PIG           & STR           & DaG           & RS            & Diag          & \color{red}{AVG}           \\ \hline
			Incep-com      & 70.9          & 87.1          & 79.7          & 66.1          & 74.2          & 60            & 77.2          & 74.2          & 73.7          \\
			HcCNN          & 70.6          & 87.1          & \textbf{84.8} & 68.6          & 71.6          & \textbf{65.6} & 80.8          & 69.9          & 74.9          \\
			FM4-FS & 70.9          & 86.8          & 81.8          & \textit{72.4} & 74.4          & 61            & \textbf{83}   & 74.9          & 75.7          \\
			CAFNet         & 70.1          & 87.8          & 84.3          & \textbf{73.4} & \textit{77}   & 61.5          & \textit{81.8} & \textbf{78.2} & \textit{76.8} \\
			TFormer        & 70.9          & 86.4          & 83.5          & 68.8          & 74            & 64.9          & 81.3          & 73            & 75.3          \\
			\textbf{AMMFM}    & \textbf{72.7} & \textbf{89.1} & 82.3          & \textit{72.4} & \textbf{78.7} & \textbf{65.8} & 81            & \textit{75.2} & \textbf{77.2} \\ \hline
	\end{tabular}}
	\vspace{-0.1in}
	\label{tab3}
\end{table}

% \usepackage{multirow}
% \usepackage{multirow}
\begin{table}[t]
	\caption{Comparisons between our proposed method AMMFM and other state-of-the-art methods in melanoma-related features ($\%$).}
	\centering
	\tabcolsep=0.5mm
	\renewcommand\arraystretch{1.2}
	\scalebox{0.6}{
		\begin{tabular}{ccccccccccc}
			\hline
			\multirow{2}{*}{Metric} & \multirow{2}{*}{Method} & DIAG & PN   & STR  & PIG  & RS   & DaG  & BWV  & VS   & \multirow{2}{*}{\color{red}{AVG}} \\
			&                         & MEL  & ATP  & IR   & IR   & PRS  & IR   & PRS  & IR   &                      \\ \hline
			\multirow{6}{*}{AUC}    & Incep-com               & 86.3 & 79.9 & 78.9 & 79   & 82.9 & 79.9 & 89.2 & 76.1 & 81.5                 \\
			& HcCNN                   & 85.6 & 78.3 & 77.6 & 81.3 & 81.9 & 82.6 & 89.8 & 82.7 & 82.5                 \\
			& FM-FS                   & 89   & 83.9 & 81.4 & 83.5 & 81.7 & 80.1 & 90.6 & 78.9 & 83.7                 \\
			& GIIN                    & 87.6 & 87.5 & 81.2 & 83.6 & 79   & 83.1 & 90.8 & 75.4 & 83.5                 \\
			& CAFNet                  & 92.2 & 75.3 & 85.4 & 85   & 85.4 & 78.7 & 94.6 & 83.4 & 85                   \\
			& \textbf{AMMFM}                   & 89.3 & 86.7 & 85.2 & 86.2 & 83.6 & 86.0 & 94.1 & 82.4 & \textbf{86.7}        \\ \hline
			\multirow{6}{*}{PRE}    & Incep-com               & 65.3 & 61.6 & 52.7 & 57.8 & 56.5 & 70.5 & 63   & 30.8 & 57.3                 \\
			& HcCNN                   & 62.8 & 62.3 & 52.4 & 65.1 & 81.6 & 69.6 & 91.9 & 50   & 67                   \\
			& FM-FS                   & 65.7 & 82.2 & 56.2 & 67.6 & 82   & 67.2 & 64.9 & 42.9 & 68.5                 \\
			& GIIN                    & 65.6 & 48.4 & 50.4 & 82.3 & 73.5 & 74.9 & 67.4 & 100  & \textbf{70.3}        \\
			& CAFNet                  & 77.9 & 50.8 & 54.8 & 70.1 & 76.7 & 67.8 & 75.4 & 58.3 & 66.5                 \\
			& \textbf{AMMFM}                  & 59.4 & 57.0 & 56.4 & 63.7 & 51.9 & 77.4 & 66.7 & 16.7 & 56.1                \\ \hline
			\multirow{6}{*}{SEN}    & Incep-com               & 61.4 & 48.4 & 51.1 & 59.7 & 66   & 62.1 & 77.3 & 13.3 & 54.9                 \\
			& HcCNN                   & 58.4 & 40.9 & 35.1 & 55.7 & 95.2 & 80.2 & 92.2 & 20   & 59.7                 \\
			& FM-FS                   & 62.4 & 49.5 & 47.9 & 58.9 & 47.1 & 68.4 & 66.7 & 20   & 52.6                 \\
			& GIIN                    & 59   & 77.5 & 67   & 39.2 & 21.9 & 70.1 & 69.9 & 3.6  & 51                   \\
			& CAFNet                  & 75.3 & 65.9 & 67.1 & 60.3 & 42.7 & 74.1 & 68.8 & 45   & 62.4                 \\
			& \textbf{AMMFM}                    & 71.4 & 63.1 & 64.6 & 71.8 & 69.6 & 74.9 & 73.5 & 31.3 & \textbf{65}          \\ \hline
			\multirow{6}{*}{SPE}    & Incep-com               & 88.8 & 90.7 & 85.7 & 80.1 & 81.3 & 78.9 & 89.4 & 97.5 & 86.6                 \\
			& HcCNN                   & 88.1 & 92.4 & 90   & 86.3 & 41.5 & 71.6 & 65.3 & 98.4 & 79.2                 \\
			& FM-FS                   & 88.8 & 90.1 & 88.4 & 88.1 & 96.2 & 72.9 & 91.6 & 97.8 & 89.2                 \\
			& GIIN                    & 89.5 & 79   & 80.3 & 95.8 & 96.8 & 78.8 & 91   & 100  & 88.9                 \\
			& CAFNet                  & 93.6 & 90.2 & 91.2 & 89.1 & 96.5 & 74   & 95.1 & 98.7 & \textbf{91.1}        \\
			& \textbf{AMMFM}                    & 86.8 & 87.1 & 86.9 & 84.2 & 83.9 & 81.1 & 92.4 & 93.4 & 87.0                 \\ \hline
	\end{tabular}}
	\label{tab4}
	\vspace{-0.1in}
\end{table}

\subsection{Comparisons with State-of-the-art Methods}

In Tables~\ref{tab2} and~\ref{tab3}, a comparative analysis was conducted to assess the performance of the proposed (AMMFM) against state-of-the-art classification methodologies utilizing clinical and dermoscopy images. 
The evaluated methods encompass Inception-combined \citep{kawahara2018seven}, HcCNN \citep{bi2020multi}, FusionM4Net-FS \citep{tang2022fusionm4net}, GIIN \citep{fu2022graph}, CAFNet \citep{he2023co}, and TFormer \citep{zhang2023tformer}.

In Table~\ref{tab2}, a comparison followed by \citep{bi2020multi, fu2022graph} was adopted to compare selected features, gauged by the AUC for each method. Table~\ref{tab3} presents a comprehensive comparison of all methods in terms of accuracy.
Note that results for all other methods were quoted from their respective publications or sourced from \citep{he2023co}. 
These results are assumed to represent the best performance, except for TFormer, where mean values and standard deviation were reported. 
Consequently, the comparison between AMMFM and TFormer is based on mean values, while comparisons with other methods are grounded on the highest value.

As shown in Table~\ref{tab2}, AMMFM attains the highest performance, boasting an averaged (AVG) AUC of $88.1\%$. 
This outperforms the second-best method, FM-FS, with $86\%$, and the third-best method, CAFNet, with $85.8\%$, by $2.1\%$ and $2.3\%$, respectively. 
AMMFM demonstrates superiority by achieving the highest values in seven categories and the second-highest in eight categories, showcasing its excellence across all eight classification tasks. 
Notably, AMMFM outperforms FM-FS in the Diag task and most Seven-Point features tasks. 
In comparison to CAFNet, AMMFM achieves comparable performance in the Diag task (AVG AUC: CAFNet: $93.1\%$, AMMFM: $92.6\%$) and clearly better performance in other Seven-Point feature tasks (AVG AUC: CAFNet: $83.0\%$, AMMFM: $86.4\%$), establishing its overall superiority over CAFNet.

Similarly, in Table~\ref{tab3}, AMMFM secures the highest averaged accuracy (AVG Acc) value of $77.2\%$, outperforming all other methods. 
It attains the highest values in four classification tasks (PN, BWV, STR, DaG) and the second-highest values in two tasks (PIG, Diag). 
CAFNet and FM-FS secure the second-best and third-best AVG Acc in Table~\ref{tab3}, respectively. 
These results underscore not only the superior performance of AMMFM, but also the efficacy of the cross-attention modules in CAFNet and the late fusion scheme in FM-FS.
For further insights, Table~\ref{tab4} details that AMMFM achieves the highest AUC and sensitivity values in melanoma-related features, substantiating its proficiency in melanoma detection. 
However, AMMFM achieves the lowest precision value, which is because of the bad performance in the categories of RS-PRE (51.9 $\%$) and VS-IR (16.7$\%$). Especially, VS-IR is much lower than other methods. 
We attribute this to the unbalanced categories in RS-PR (ABS: 758 PRE: 253) and VS-IR (ABS: 833, REG: 117 and IR: 91, see Table~\ref{table1}).

\begin{table}[t]
	\caption{Comprehensive comparison between our proposed method AMMFM and other state-of-the-art methods in terms of model parameters.
		For the approximated model parameter, $>$ denotes slightly more, $\gg$ denotes much more.
	}
	\centering
	\tabcolsep=1mm
	\renewcommand\arraystretch{1.0}
	\scalebox{1}{
		\begin{tabular}{c|ccc}
			\hline
			Method         & AVG AUC       & AVG ACC       & \color{red}{Parameters (Mb)}     \\ \hline
			Incep-com      & 83.7          & 73.7          & $>$ 57,4  \\
			HcCNN          & 84.3          & 74.9          & $\gg$ 51.2  \\
			FM-FS          & 86            & 75.7          & 54.45               \\
			GIIN           & 83.6          & -             & $>$ 51.2  \\
			CAFNet         & 85.8          & 76.8          & $\gg$ 51.2  \\
			TFormer        & -             & 75.3          & 77.76               \\
			\textbf{AMMFM} & \textbf{88.1} & \textbf{77.2} & \textbf{33.06}      \\ \hline
	\end{tabular}}
	\label{tab5}
	\vspace{-0.1in}
\end{table}
In Table~\ref{tab5}, we compare all the methods mainly based on model parameters, and for convenience, we again present the AVG AUC and AVG ACC in this table.
Since only the source codes for FM-FS and TFormer were available, the model parameters for Incep-com, HcCNN, GIIN, and CAFNet were estimated by us, albeit with some approximations.
In our estimation, we calculated the parameters based on the employed backbones, specifically, two InceptionV3 (57.4Mb) for Incep-com and two ResNet-50 (51.2Mb) for GIIN, HcCNN, and CAFNet.
However, concerning Incep-com and GIIN, the model parameters are marginally higher than those of their utilized backbones, attributed to the absence of multiple blocks employed in constructing the third branch.
Concerning HcCNN and CAFnet, which construct a third branch utilizing attention and ResNet blocks, the model parameters are expected to significantly exceed those of their backbones. 
As shown in Table.~\ref{tab5}, our AMMFM achieves the highest values in both AVG AUC and AVG ACC with the least model parameters (33.06Mb), demonstrating the great accuracy/parameter trade-off of our AMMFM.

%and the effectiveness of our proposed asymmetrical fusion framework and asymmetrical attention blocks
% \usepackage{multirow}

\subsection{Ablation Studies}
In following experiments, all the models are trained and tested ten times to obtain the mean value and standard deviation for a fair comparison.

In Table~\ref{tab6}, ablation studies are conducted to analyze the two primary components of our AMMFM: the asymmetrical fusion framework (AFF) and the asymmetrical attention block (AAB). 
For comparative purposes, we establish a baseline utilizing a commonly-used symmetrical fusion framework (SFF) based on two Swin-Transformer (ST) models and a concatenation operation, serving as a reference point in the ablation studies.
\begin{table}[t]
	\centering
	\caption{Ablation studies of our AMMFM in terms of AVG AUC, AVG ACC and model parameters. ($\%$)}
	\tabcolsep=1mm
	\renewcommand\arraystretch{1.2}
	\scalebox{1}{
		\begin{tabular}{ccccc}
			\hline
			AFF           & AAB          & AVG AUC    & AVG ACC    & Parameters(Mb) \\ \hline
			\multicolumn{2}{c}{Baseline} & 87.2$\pm$0.4 & 76.6$\pm$0.5  & 58.49         \\
			$\checkmark$             &              & 87.2$\pm$0.4 & 76.5$\pm$0.4 & 32.48         \\
			$\checkmark$              & $\checkmark$             & 87.6$\pm$0.4 & 76.7$\pm$0.4 & 33.06         \\ \hline
	\end{tabular}}
	\label{tab6}
	\vspace{-0.1in}
\end{table}

Compared to baseline model (second row), the proposed AFF with concatenation operation (third row) can significantly reduce the parameters from 58.49M to 32.48M without compromising the performance metrics, as evidenced by the maintained the AVG AUC (Baseline: 87.2$\%$, AFF: 87.2$\%$) and AVG ACC (Baseline:76.6$\%$, AFF: 76.5$\%$). 
These outcomes substantiate our initial hypothesis that substituting an advanced model (ST) with a more lightweight model, MobileNetV3 (MN), for information extraction from clinical images, would have a subtle or negligible impact on overall performance. 
This observation underscores the supplementary nature of clinical images in the context of multi-label skin lesion classification tasks, where dermoscopy images remain the primary source of information.
Subsequently, ABB contributes to a further enhancement in the performance of AFF from. 
This improvement is observed across both metrics, with AVG AUC increasing from 87.2$\%$ for AFF to 87.6$\%$ for AFF+AAB, and AVG ACC from 76.5$\%$ to 76.7$\%$. Remarkably, this performance boost is achieved with only a marginal increase in model size, rising from 32.48Mb for AFF to 33.06Mb for AFF+AAB, illuminating the effectiveness of our AAB in accuracy/parameters trade-off.

\begin{table}[h]
	\caption{Comparison between single-modal, baseline multi-modal and our proposed multi-modal methods. Clin: Clinical Images, Derm: Dermoscopy Images, MN: MobilenetV3, ST: Swin-Transformer, Param: Parameters and - indicates no model for extracting the information from dermoscopy and clinical images, i.e., single-modal methods. ($\%$)}
	\centering
	\tabcolsep=1mm
	\renewcommand\arraystretch{1.2}
	\scalebox{0.92}{
		\begin{tabular}{cccccc}
			\hline
			Method                        & Clin & Derm & AVG AUC   & AVG ACC   & Params(Mb)             \\ \hline
			\multirow{4}{*}{Single-Modal} & -    & ST   & 86.6$\pm$0.3 & 76.2$\pm$0.6 & \multirow{2}{*}{28.82} \\
			& ST   & -    & 77.3$\pm$0.5 & 68.7$\pm$0.7 &                        \\
			& -    & MN   & 83.1$\pm$0.4 & 72.6$\pm$0.3 & \multirow{2}{*}{2.91}  \\
			& MN   & -    & 75.7$\pm$0.4 & 67.6$\pm$0.5 &                        \\ \hline
			\multirow{2}{*}{Baseline}     & ST   & ST   & 87.2$\pm$0.4 & 76.6$\pm$0.5 & 58.49                  \\
			& MN   & MN   & 84.5$\pm$0.3 & 73.4$\pm$0.2 & 6.48                   \\ \hline
			\multirow{2}{*}{AMMFM}        & ST   & MN   & 84.7$\pm$0.3 & 73.8$\pm$0.3 & 32.62                  \\
			& MN   & ST   & 87.6$\pm$0.4 & 76.7$\pm$0.4 & 33.06                  \\ \hline
		\end{tabular}
	}
	\label{tab7}
	\vspace{-0.1in}
\end{table}

\subsection{Comparison between single-modal, baseline multi-modal and our proposed multi-modal methods}
In Table~\ref{tab7}, we conduct a comprehensive comparison among single-modal, baseline multi-modal (SFF with concatenation), and our proposed multi-modal approach (AFF with AAB).
The results reveal a substantial performance improvement when utilizing dermoscopy images, regardless of whether based on MN or ST, compared to clinical images. 
Specifically, there is an increase of over 7$\%$ in AVG AUC and 5$\%$ in AVG ACC values, underscoring the pronounced significance of dermoscopy images in the context of multi-label classification tasks.
Furthermore, the baseline multi-modal methods exhibit an additional increase in accuracy compared to their Derm-based counterparts. 
This emphasizes the complementary nature of clinical images, which provide supplementary information to dermoscopy images.
In the single-modal approaches, substituting ST with MN for dermoscopy image processing results in a 3.5$\%$ reduction in both metrics. 
Conversely, replacing ST with MN for clinical image processing shows a more modest 1.1$\%$ decrease in Average Accuracy (AVG ACC).
Also, compared to another counterpart in AMMFM approaches, employing ST for the dermoscopy branch and MN for the clinical branch enhances both metrics by approximately 3$\%$. 
These results demonstrate support the effectiveness of our AFF and affirm the suitability of the chosen architecture for efficiently optimizing performance in multi-modal classification tasks.

In Table~\ref{tab8}, we present detailed information about both ST-based single-modal and multi-modal methods, facilitating a nuanced analysis of their impact on individual classification tasks.
Examining the table reveals that dermoscopy-based ST consistently outperforms clinical image-based ST across all categories (CTs), a result aligned with expectations given that the seven-point checklist criteria are formulated based on observed features under dermoscopy \citep{kawahara2018seven}. 
Moreover, in comparison to dermoscopy-based ST, both the Baseline and our AMMFM demonstrate performance improvements across nearly all CTs. 
This observation illustrates the complementary role of clinical images in enhancing the overall performance when combined with dermoscopy images.
\begin{table}[t]
	\centering
	\caption{Detailed comparison between ST-based single-modal and multi-modal in terms of AVG AUC. CT: classification task, CG: category, Clin: clinical images, Derm: Dermscopy Images. $(\%)$}
	\tabcolsep=1mm
	\renewcommand\arraystretch{1.2}
	\scalebox{0.8}{
		\begin{tabular}{cclccc}
			\hline
			CT                    & CG   & \multicolumn{1}{c}{\makecell[c]{Clin \\ (ST)}} & {\makecell[c]{Derm \\ (ST)}}      & {\makecell[c]{Baseline \\ (two STs)}}           & {\makecell[c]{AMMFM \\ (Clic: MN \\ Derm: ST)}}                 \\ \hline
			\multirow{5}{*}{Diag} & BCC  & 85.5$\pm$3.8                & 94.5$\pm$1.3 & 95.0$\pm$0.9          & \textbf{95.0$\pm$0.8} \\
			& NEV  & 85.8$\pm$0.7                & 91.8$\pm$0.5 & \textbf{92.4$\pm$0.4} & \textbf{92.4$\pm$0.4} \\
			& MEL  & 79.5$\pm$1.1                & 89.1$\pm$0.6 & 89.3$\pm$0.8          & \textbf{89.5$\pm$0.7} \\
			& MISC & 86.5$\pm$1.8                & 93.7$\pm$0.8 & 93.9$\pm$0.9          & \textbf{94.7$\pm$0.7} \\
			& SK   & 74.5$\pm$4.2                & 87.5$\pm$3.2 & 88.3$\pm$2.0          & \textbf{90.4$\pm$1.4} \\
			\multirow{2}{*}{PN}   & TYP  & 80.4$\pm$1.1                & 87.5$\pm$0.6 & \textbf{88.1$\pm$0.7} & 87.7$\pm$0.8          \\
			& ATP  & 73.4$\pm$1.6                & 85.1$\pm$0.7 & 85.7$\pm$0.7          & \textbf{86.2$\pm$0.8} \\
			\multirow{2}{*}{STR}  & REG  & 79.9$\pm$2.2                & 89.3$\pm$1.1 & 88.6$\pm$1.3          & \textbf{89.4$\pm$1.8} \\
			& IR   & 71.0$\pm$1.6                & 83.9$\pm$1.0 & 84.2$\pm$0.9          & \textbf{84.6$\pm$1.1} \\
			\multirow{2}{*}{PIG}  & REG  & 68.3$\pm$2.1                & 82.2$\pm$1.5 & 81.4$\pm$1.4          & \textbf{83.5$\pm$1.1} \\
			& IR   & 73.6$\pm$1.7                & 85.5$\pm$1.3 & \textbf{85.5$\pm$0.9} & 85.4$\pm$1.1          \\
			RS                    & PRS  & 72.9$\pm$1.3                & 82.5$\pm$0.9 & \textbf{83.5$\pm$0.9} & 83.4$\pm$0.8          \\
			\multirow{2}{*}{DaG}  & REG  & 72.6$\pm$1.4                & 79.3$\pm$1.0 & 80.2$\pm$0.9          & \textbf{80.3$\pm$0.6} \\
			& IR   & 73.8$\pm$0.9                & 84.1$\pm$0.8 & 84.1$\pm$0.9          & \textbf{84.4$\pm$0.7} \\
			BWV                   & PRS  & 83.9$\pm$1.2                & 92.7$\pm$0.7 & \textbf{93.8$\pm$0.5}          & 93.8$\pm$0.8          \\
			\multirow{2}{*}{VS}   & REG  & 81.7$\pm$2.0                & 86.2$\pm$1.1 & \textbf{88.3$\pm$0.8} & 87.9$\pm$1.3          \\
			& IR   & 77.0$\pm$2.2                & 80.4$\pm$1.7 & \textbf{82.2$\pm$1.5} & 81.3$\pm$2.3          \\ \hline
			\multicolumn{2}{c}{AVG}      & 77.3$\pm$0.5                & 86.6$\pm$0.3 & 87.2$\pm$0.4          & \textbf{87.6$\pm$0.4} \\ \hline
	\end{tabular}}
	\label{tab8}
	\vspace{-0.1in}
\end{table}

\subsection{Comparison between bidirectional attention block (BAB) and asymmetrical attention block (AAB)}
% \usepackage{multirow}
\begin{table}[t]
	\caption{The comparison between our proposed asymmetrical attention block (AAB) and other fusion blocks (FBs), including concatenation (CAT) and bidirectional attention block (BAN), in different fusion frameworks (FS). SFF: symmetrical fusion framework, AFF: asymmetrical fusion framework ($\%$).}
	\centering
	\tabcolsep=1mm
	\renewcommand\arraystretch{1.2}
	\scalebox{1}{
		\begin{tabular}{c|cccccc}
			\hline
			\multirow{2}{*}{FS}  & \multicolumn{3}{c}{FB} & \multirow{2}{*}{AVG AUC} & \multirow{2}{*}{AVG ACC} & \multirow{2}{*}{Params(Mb)} \\ \cline{2-4}
			& CAT    & BAB   & AAB   &                          &                          &                             \\ \hline
			\multirow{3}{*}{SFF} & $\checkmark$      &       &       & 87.2$\pm$0.4                & 76.6$\pm$0.5                & 58.49M                      \\
			&        & $\checkmark$     &       & 87.2$\pm$0.3                & 76.6$\pm$0.4                & 60.45M                      \\
			&        &       & $\checkmark$     & \textbf{87.5$\pm$0.4}                & \textbf{77.1$\pm$0.4}                & 59.09M                      \\ \hline
			\multirow{3}{*}{AFF} & $\checkmark$      &       &       & 87.2$\pm$0.4                & 76.5$\pm$0.4                & 32.48M                      \\
			&        & $\checkmark$     &       & 87.5$\pm$0.3                & 76.6$\pm$0.3                & 33.88M                      \\
			&        &       &$\checkmark$     & \textbf{87.6$\pm$0.4}                & \textbf{76.7$\pm$0.4}                & 33.06M                      \\ \hline
	\end{tabular}}
	\label{tab9}
	\vspace{-0.1in}
\end{table}

To delve deeper into the impact of the proposed asymmetrical attention block (AAB), a comparative analysis is conducted with other fusion blocks within different fusion frameworks, namely the symmetrical fusion framework (SFF) and the asymmetrical fusion framework (AFF).
As illustrated in Table~\ref{tab9}, the block attention block (BAB) demonstrates performance improvement over concatenation (CAT) for both fusion frameworks, highlighting the effectiveness of multi-modal interactions. 
Notably, our proposed AAB further enhances the accuracy achieved by BAB within both SFF and AFF. 
This observation lends support to our assumption that over augmenting the importance of clinical supplementary information in the multi-modal pipeline may impact the classification task. 
At the same time, ABB also show it's superiority in model's parameters compared to BAB.

\begin{table}[h]
	\caption{Comparison between single-modal, baseline multi-modal and our AMMFM methods based on different backbones. ST: Swin-Transformer, Params: Parameters ($\%$).}
	\centering
	\tabcolsep=1mm
	\renewcommand\arraystretch{1.2}
	\scalebox{1}{
		\begin{tabular}{c|cccc}
			\hline
			Backbone                  & Modal          & AVG AUC            & AVG ACC            & Params(Mb)             \\ \hline
			\multirow{4}{*}{ResNet}       & Derm           & 84.0$\pm$0.5          & 74.0$\pm$0.4          & \multirow{2}{*}{26.68} \\
			& Clin           & 76.6$\pm$0.5          & 67.5$\pm$0.4          &                        \\
			& Baseline       & 85.4$\pm$0.2          & 74.6$\pm$0.4          & 55.52                  \\
			& \textbf{AMMFM} & \textbf{85.7$\pm$0.2} & \textbf{74.9$\pm$0.3} & 36.36                  \\ \hline
			\multirow{4}{*}{ConvNext} & Derm           & 86.5$\pm$0.4          & 75.9$\pm$0.3          & \multirow{2}{*}{29.05} \\
			& Clin           & 77.5$\pm$0.3          & 69.0$\pm$0.3          &                        \\
			& Baseline       & 87.1$\pm$0.3          & 76.5$\pm$0.3          & 58.96                  \\
			& \textbf{AMMFM} & \textbf{87.3$\pm$0.5} & \textbf{76.7$\pm$0.4} & 33.29                  \\ \hline
			\multirow{4}{*}{ST}       & Derm           & 86.6$\pm$0.3          & 76.2$\pm$0.6          & \multirow{2}{*}{28.82} \\
			& Clin           & 77.3$\pm$0.5          & 68.7$\pm$0.7          &                        \\
			& Baseline       & 87.2$\pm$0.4          & 76.6$\pm$0.5          & 58.49                  \\
			& \textbf{AMMFM} & \textbf{87.6$\pm$0.4} & \textbf{76.7$\pm$0.4} & 33.06                  \\ \hline
	\end{tabular}}
	\label{tab10}
	\vspace{-0.1in}
\end{table}

\subsection{Generalization ability of AMMFM using different backbones}
To assess the generalization capability of our AMMFM across various backbones, we employ ResNet-50, ConvnextTiny, and Swin-Transformer as backbone architectures for comparative analysis.
As depicted in Table~\ref{tab10}, our AMMFM consistently outperforms the baseline multi-modal method across all three backbones, while maintaining significantly fewer parameters. 
This substantiates the robustness of our AMMFM, showcasing its ability to deliver superior performance across diverse deep learning backbones.

\section{Conclusion}
In this study, we introduced a novel Asymmetrical Multi-Modal Fusion Method (AMMFM) for efficient multi-label skin lesion classification, driven by the observation that dermoscopy images provide more crucial information than clinical images. 
Our AMMFM comprises two key components: the asymmetrical fusion framework (AFF) and the asymmetrical attention block (AAB).
To optimize efficiency by reducing parameters, AFF integrates one advanced model for feature extraction from dermoscopy images and one lightweight model for clinical images. 
This design is grounded in the assumption that affecting the ability to capture supplementary information from clinical images will subtly or not impact the overall multi-modal pipeline's performance.
In contrast to the bidirectional attention block (BAB), AAB focuses solely on enhancing dermoscopy features while excluding attention on clinical images, due to our belief that directing attention to supplementary information may adversely impact the final classification performance.
Extensive results demonstrate that, in comparison to the previous symmetrical fusion framework, AFF significantly reduces model parameters while maintaining accuracy. 
Additionally, AAB enhances the performance of BAB with fewer parameters, showcasing its efficacy in improving the overall classification task.
Last but not least, our AMMFM attains state-of-the-art performance with the fewest model parameters.

	\bibliographystyle{elsarticle-harv} 
	\bibliography{myreference_11_08}
	%\end{thebibliography}
\end{document}